\documentclass[10pt, letterpaper]{article}

\usepackage[margin=1in]{geometry}
\usepackage{amsmath, amsthm, amssymb}
\usepackage{hyperref}

\usepackage[labelfont=bf, tableposition=top]{caption}
\usepackage{graphicx, color}

\usepackage[indent=40pt]{parskip}
\title{Persistent homology analysis of type 2 diabetes genome-wide association studies in protein–protein interaction networks}

\author{Euijun Song\thanks{Yonsei University College of Medicine, Seoul, Republic of Korea; Present: Independent Researcher, Gyeonggi, Republic of Korea; \href{mailto:drjunsong@gmail.com}{\texttt{drjunsong@gmail.com}}}}

\date{\today}

\begin{document}

\maketitle

\begin{abstract}
Genome-wide association studies (GWAS) involving increasing sample sizes have identified hundreds of genetic variants associated with complex diseases, such as type 2 diabetes (T2D); however, it is unclear how GWAS hits form unique topological structures in protein–protein interaction (PPI) networks. Using persistent homology, this study explores the evolution and persistence of the topological features of T2D GWAS hits in the PPI network with increasing P-value thresholds. We define an $n$-dimensional persistent disease module as a higher-order generalization of the largest connected component (LCC). The 0-dimensional persistent T2D disease module is the LCC of the T2D GWAS hits, which is significantly detected in the PPI network (196 nodes and 235 edges, P$<$0.05). In the 1-dimensional homology group analysis, all 18 1-dimensional holes (loops) of the T2D GWAS hits persist over all P-value thresholds. The 1-dimensional persistent T2D disease module comprising these 18 persistent 1-dimensional holes is significantly larger than that expected by chance (59 nodes and 83 edges, P$<$0.001), indicating a significant topological structure in the PPI network. Our computational topology framework potentially possesses broad applicability to other complex phenotypes in identifying topological features that play an important role in disease pathobiology.
\end{abstract}

\section{Introduction}
Understanding the genotype–phenotype relationships is challenging owing to their polygenicity and nonlinearity. Complex diseases result from interactions between diverse cellular processes and genes. Elucidating the genetic basis of complex diseases in the context of protein–protein interaction (PPI) networks is essential \cite{Barabasi2011, BarrioHernandez2023}. In the PPI network, genes (or gene products) that have similar biological functions are likely to interact closely with each other. Thus, genes associated with a specific phenotype tend to be clustered into a connected component called a \textit{disease module} in the PPI network \cite{Goh2007}. Disease modules that significantly overlap with each other exhibit similar pathobiological pathways, co-expression patterns, and clinical manifestations \cite{Menche2015}. This disease module concept is useful in identifying novel disease–disease or disease–drug relationships \cite{Menche2015, Guney2016}, enabling the implementation of network-based drug repurposing for complex traits/diseases \cite{Song2020}.

Genome-wide association studies (GWAS) have identified numerous genetic variants associated with various complex diseases and can be used to characterize disease-associated modules in the PPI network \cite{BarrioHernandez2023}. Genes associated with GWAS loci or GWAS hits tend to be mapped onto coherent network modules in the PPI network. As the P-value threshold increases from 0 to the standard genome-wide significance threshold of 5×10$^{-8}$, the GWAS hits mapped to the PPI network tend to gradually form a single, connected component \cite{Menche2015, Leopold2018}. This largest connected component (LCC) of disease genes is occasionally called an \textit{observable disease module} \cite{Menche2015, Paci2021}. However, owing to the limited sample size and coverage of current GWAS data as well as the interactome incompleteness, disease-associated seed genes are often scattered in the PPI network. To detect disease modules, various seed-expanding and/or heuristic-based algorithms have been developed to expand and merge the scattered seed genes in the PPI network \cite{Choobdar2019, Ghiassian2015, Vlaic2018, Wang2018}. In addition, deep learning and graph neural network models have been used to predict disease-associated genes \cite{Hou2022, Yang2019, Yang2022} and disease treatment mechanisms \cite{Ruiz2021, Zitnik2018} in the context of biological networks. Most studies have focused on identifying connected components by mapping disease-associated seed genes onto the PPI network and expanding these seed genes. However, the mechanism by which GWAS hits are mapped onto the LCC or other unique topological structures in the PPI network as the P-value threshold increases remains unclear. Therefore, the topological features of GWAS hits mapped onto the PPI network warrant investigation.

One mathematical method for analyzing the topological features of complex networks is simplicial homology \cite{Hatcher2002, Salnikov2019}. Simplicial homology is an algebraic topology tool used to analyze the topological features of a simplicial complex, which is a collection of higher-order interactions called simplices, including points (0-simplices), line segments (1-simplices), triangles (2-simplices), and higher-dimensional simplices. Simplicial homology can be used to examine the connectivity patterns within biological networks, such as gene-regulatory networks or brain connectivity networks. It can identify topological features, such as connected components (0-holes), loops (1-holes), voids (2-holes), and higher-dimensional holes in the data. For example, the LCC is the largest 0th homology class (connected component). Persistent homology is a method for capturing the persistence of simplicial homology features across multiple thresholds corresponding to a filtration of the simplicial complex \cite{Salnikov2019, Otter2017}. It can identify important topological features that are persistent across different levels of interaction, rather than artifacts of noise or parameter uncertainty. Persistent homology features of biological networks potentially correspond to biologically relevant components that play a crucial role in disease mechanisms \cite{Masoomy2021, Sizemore2019}. The mathematical details of simplicial complex and homology concepts are described in the Method section.

This study analyzes the persistent homology features of GWAS hits in the PPI network to identify important topological structures that potentially play a significant role in disease pathobiology. We analyze the simplicial homology features of GWAS hits in the PPI network as the P-value threshold increases from 0 to 5×10$^{-8}$. For example, the LCC of the mapped GWAS hits, which is occasionally called an observable disease module, can be considered a connected component (0th homology class) that lives forever. This study aims to expand the LCC concept using higher-order topological structure analysis. We use GWAS summary statistics data of type 2 diabetes (T2D) because T2D has undergone extensive genetic study across diverse ancestry populations with large sample sizes. GWAS with increasing sample sizes have recently identified more than 300 genetic loci associated with T2D \cite{Vujkovic2020}; however, many of these GWAS loci have small effect sizes of unclear pathobiological meaning. Therefore, this study systematically explores the evolution and persistence of the topological features of T2D GWAS hits in the PPI network as the P-value threshold increases from 0 to 5×10$^{-8}$. We also analyze biological pathways, transcription factors, and microRNAs associated with the persistent homology features.

\section{Methods}

\subsection{Overview of the computational topology framework}
This study analyzes the topological features of GWAS hits in the human PPI network. Using persistent homology, we systematically explore $n$-dimensional holes associated with a specific phenotype, as follows:

\begin{enumerate}
\item Map GWAS hits onto the human PPI network.
\item Using persistent homology, identify $n$-dimensional holes of GWAS hits in the PPI network, as the P-value threshold increases from 0 to 5×10$^{-8}$.
\item Detect \textit{$n$-th persistent disease modules}, which we define as unions of $n$-dimensional holes that live forever over all P-value thresholds.
\item Compute the statistical significance of $n$-th persistent disease modules by comparing the result with the randomized distribution of a set of randomly selected nodes in the PPI network.
\end{enumerate}

Since the LCC can be considered a connected component (0th homology class) that lives forever, the $n$-th persistent disease module can be viewed as a higher-order generalization of the LCC concept. We test our computational framework using T2D GWAS summary statistics data, and perform functional enrichment analysis to validate the pathobiological significance of the persistent homology features.

\subsection{Consolidated human protein–protein interactome}
We used a consolidated human PPI network constructed previously by Wang and Loscalzo \cite{Wang2021, Wang2023}. Briefly, the protein–protein interactome was compiled from various sources, including high-throughput yeast-two-hybrid studies, the Center for Cancer Systems Biology (CCSB) human interactome, binary PPIs from other laboratories, protein–protein co-complex interactions, signaling interactions, kinase–substrate interactions, and the Human Reference Interactome (HuRI) binary PPIs. This network possesses a scale-free topology \cite{Wang2021}. The LCC of the protein–protein interactome, comprising 16,422 proteins (nodes) and 233,940 interactions (links), was used for the downstream analyses.

\subsection{T2D GWAS hits}
We used a GWAS meta-analysis summary statistics dataset of 228,499 T2D cases and 1,178,783 controls encompassing multi-ancestral groups \cite{Vujkovic2020} (downloaded from the GWAS catalog \\ https://www.ebi.ac.uk/gwas/). The standard genome-wide significance threshold of 5×10$^{-8}$ was applied. Each genetic variant was annotated with the closest gene(s) via GWAS catalog gene-mapping data. Only GWAS loci that had been annotated with at most two genes were included. Some genes were linked to multiple GWAS loci with multiple P-values. To extract GWAS hits, for each gene, we assigned the lowest P-value from the different GWAS loci mapped onto that gene \cite{Ratnakumar2020}. We only considered genes (or proteins) in the human PPI network.

\subsection{Simplicial complex and homology theory}
Here, we briefly describe the fundamentals of the simplicial complex and persistent homology theory \cite{Hatcher2002, Salnikov2019, Otter2017}. An $n$-dimensional simplex ($n$-simplex) is formed by $n+1$ nodes
\begin{equation}
\sigma_n=\left(v_0,v_1,\ldots,v_n\right)
\end{equation}
with an assigned orientation. For example, a 0-simplex is a vertex (node), a 1-simplex is an edge (link), and a 2-simplex is a triangle. An $n^\prime$-face of an $n$-simplex ($n^\prime<n$) is a proper subset of the nodes of the simplex with order $n^\prime+1$. A simplicial complex $K$ is a set of simplices closed under the inclusion of the faces of each simplex. Given a set of $n$-simplices of a simplicial complex $K$, an $n$-dimensional chain ($n$-chain) is defined as a finite linear combination of $n$-simplices of $K$, as follows:
\begin{equation}
c_n=\sum_{i}{b_i\sigma_n^{\left(i\right)}}
\end{equation}
where $b_i\in\mathbb{Z}/2\mathbb{Z}$. In this study, we restrict our analysis to homology with $\mathbb{Z}/2\mathbb{Z}$ coefficients. The set of $n$-chains forms an abelian group denoted by $C_n$ ($n$-chain group). For any $n$-simplex $\sigma_n=\left(v_0,v_1,\ \ldots,\ v_n\right)$, the boundary operator $\partial_n:C_n\rightarrow C_{n-1}$ is the homomorphism defined as follows:
\begin{equation}
\partial_n\left(\sigma_n\right)=\sum_{i=0}^{n}{\left(-1\right)^i\left(v_0,\ldots,v_{i-1},\ v_{i+1},\ldots,v_n\right)}.
\end{equation}
An $n$-chain is said to be a $n$-cycle if its boundary is zero; that is, elements of the subgroup $Z_n:=\ker\partial_n \subseteq C_n$ are $n$-cycles. Similarly, elements of the subgroup $B_n:={\rm{im}}\ \partial_{n+1} \subseteq C_n$ are said to be $n$-boundaries. Based on the definition of the boundary operator, it is obvious that any boundary has no boundary (i.e., $\partial_n\partial_{n+1}=0$). Thus, $B_n\subseteq Z_n\subseteq C_n$. Hence, the $n$-th simplicial homology group $H_n$ of the simplicial complex $K$ can be defined as the quotient abelian group:
\begin{equation}
H_n\left(K\right):=Z_n/B_n=\ker\partial_n/{{\rm{im}}\ \partial_{n+1}}.
\end{equation}
The rank of the $n$-th homology group $H_n$ is called the $n$-th Betti number $\beta_n$. The $n$-th homology group $H_n$ is isomorphic to $\mathbb{Z}^{\beta_n}$, with the basis of independent $n$-cycles on $Z_n$ modulo boundaries. Intuitively, it represents $n$-dimensional holes in the simplicial complex $K$. For example, $\beta_0$, $\beta_1$, and $\beta_2$ represent the number of connected components, loops, and voids, respectively.

Persistent homology is a method for analyzing simplicial topological features at different resolutions of a given simplicial complex \cite{Salnikov2019, Otter2017}. Formally, a filtration of the simplicial complex $K$ is a finite sequence of subcomplexes $\left\{K_i\ |\ 0\le i\le m\right\}$ such that
\begin{equation}
\emptyset=K_0\subseteq K_1\subseteq \cdots \subseteq K_m=K.
\end{equation}
For $0\le i\le j\le m$, the inclusion $K_i\hookrightarrow K_j$ induces a homomorphism $h_n^{i,\ j}:H_n\left(K_i\right)\rightarrow H_n\left(K_j\right)$, and the $n$-th persistent homology groups ${PH}_n^{i,\ j}$ are defined as the images of these homomorphisms:
\begin{equation}
{PH}_n^{i,\ j}:={\rm{im}}\ h_n^{i,\ j}.
\end{equation}
Intuitively, the $n$-th persistent homology groups represent $n$-dimensional holes that persist from $K_i$ to $K_j$. We can track when $n$-dimensional holes appear (birth) and disappear (death) at different threshold values of the filtration. Persistence diagrams, representations of persistent homology, can be constructed by plotting the birth and death sites of topological features.

\subsection{Persistent homology analysis of GWAS hits}
In this study, the PPI network $G=(V, E)$ is considered a simplicial complex $K$: genes (or proteins) are regarded as 0-simplexes (nodes), PPIs as 1-simplexes (links), and higher-order connections (or cliques) as high-dimensional simplices. The T2D disease module was identified as the LCC of the PPI subnetwork induced by the T2D GWAS hits. The statistical significance of the LCC was calculated by comparing the observed LCC size with the randomized LCC distribution of a set of randomly selected nodes of the same size in a degree-preserving manner over 1,000 repetitions. The $z$-score was estimated as $z=\frac{LCC_{obs}-\langle LCC\rangle_{rnd}}{\sigma_{rnd}}$, where $LCC_{obs}$ is the observed LCC size, and $\langle LCC\rangle_{rnd}$ and $\sigma_{rnd}$ are the mean and SD of the randomized LCC distribution, respectively.

Each T2D GWAS hit's P-value was used as a varying threshold to obtain a filtration of the PPI subnetwork induced by the T2D GWAS hits as a function of the P-value. As the threshold value increases from 0 to 5×10$^{-8}$, each node appears at the P-value assigned to that gene. Formally, we define the $\delta$-simplicial complex for the P-value threshold $\delta\ge 0$ as follows:
\begin{equation}
\mathcal{W}_\delta:=\left\{\sigma=\left(v_0,v_1,\ldots,v_k\right)\in K\ |\ \forall i\in\{0,1,\ldots,k\},\ p(v_{i})\le\delta\right\}
\end{equation}
where $p(v)\in[0, 1]$ is the GWAS hit P-value assigned to the node $v\in V$. Using this $\delta$-simplicial complex, we define the filtration as $\left\{\mathcal{W}_\delta \hookrightarrow \mathcal{W}_{\delta^\prime}\right\}_{0\le \delta\le \delta^\prime}$. We subsequently examined the persistent homology features ($n$-dimensional holes) of this filtration for each dimension as a function of the P-value threshold. Persistence diagrams were used to visualize the birth and death times of topological features. For each dimension, we also computed the Betti numbers (ranks) of the simplicial homology groups as a function of the P-value threshold.

We define an \textit{$n$-th persistent disease module} as a union of $n$-dimensional holes that live forever over all P-value thresholds. This definition is concordant with the conventional disease module concept – the 0th persistent disease module is the LCC, which is the persistent 0-dimensional hole (connected component) that lives forever. The statistical significance of the $n$-th persistent disease module was calculated by comparing the observed persistent disease module size with the randomized persistent disease module distribution of a set of randomly selected nodes of the same size in a degree-preserving manner. The persistent homology features of randomly selected nodes were analyzed over 1,000 repetitions. The network and homology analyses were performed using the NetworkX and Ripser packages of Python 3.8 (https://www.python.org/), and networks were visualized using Cytoscape 3.9.1 (https://cytoscape.org/). The core code for analyzing persistent homology is publicly available in our GitHub repository (https://github.com/esong0/PHGWAS).

\subsection{Functional enrichment analysis}
To infer the biological significance of the persistent disease module, a pathway enrichment analysis was performed based on the Kyoto Encyclopedia of Genes and Genomes (KEGG) 2021 database using the GSEApy Python package \cite{Fang2022} with the Enrichr web server \cite{Kuleshov2016}. In addition, the transcription factor target enrichment analysis was conducted based on the ENCODE and ChEA Consensus databases. The microRNA target enrichment analysis was also performed based on the miRTarBase database. Adjusted P-values were computed using the Benjamini–Hochberg method, and statistical significance was set at P$<$0.05.

\section{Results}

\subsection{The LCC of GWAS hits}
\begin{figure}[h]
\centering
\includegraphics[width=0.85\linewidth]{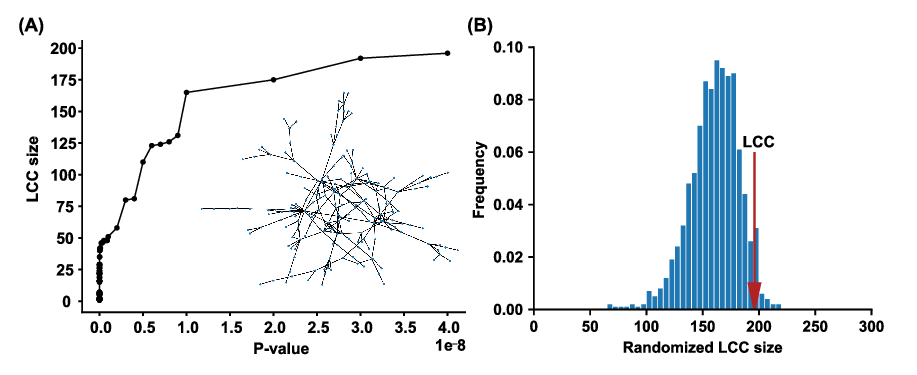}
\caption{The LCC of the subnetwork induced by the T2D GWAS hits in the protein–protein interaction network. (A) The LCC size increases as the P-value threshold increases from 0 to 5×10$^{-8}$. (B) The LCC of the T2D GWAS hits is significantly larger than that expected by chance. The red arrow indicates the observed LCC size. LCC, the largest connected component; T2D, type 2 diabetes; GWAS, genome-wide association study.}
\label{fig:fig1}
\end{figure}

We compiled the T2D GWAS hits using the large-scale T2D GWAS summary statistics data, 565 of which are present in the human PPI network. As the P-value threshold increased from 0 to 5×10$^{-8}$, the LCC of the subnetwork induced by the T2D GWAS hits increased (Figure \ref{fig:fig1}A). When the standard genome-wide significance threshold of 5×10$^{-8}$ was applied, we identified the LCC comprising 196 nodes and 235 edges, which is significantly larger than that expected by chance (P=0.0487, Figure \ref{fig:fig1}B). We defined a T2D observable disease module as this LCC of the T2D GWAS hits (Supplementary Table S1). Other connected components of the subnetwork induced by the T2D GWAS hits comprised $\le$5 nodes, which were excluded from the downstream analyses.

\subsection{Persistent homology analysis}
\begin{figure}[h]
\centering
\includegraphics[width=0.85\linewidth]{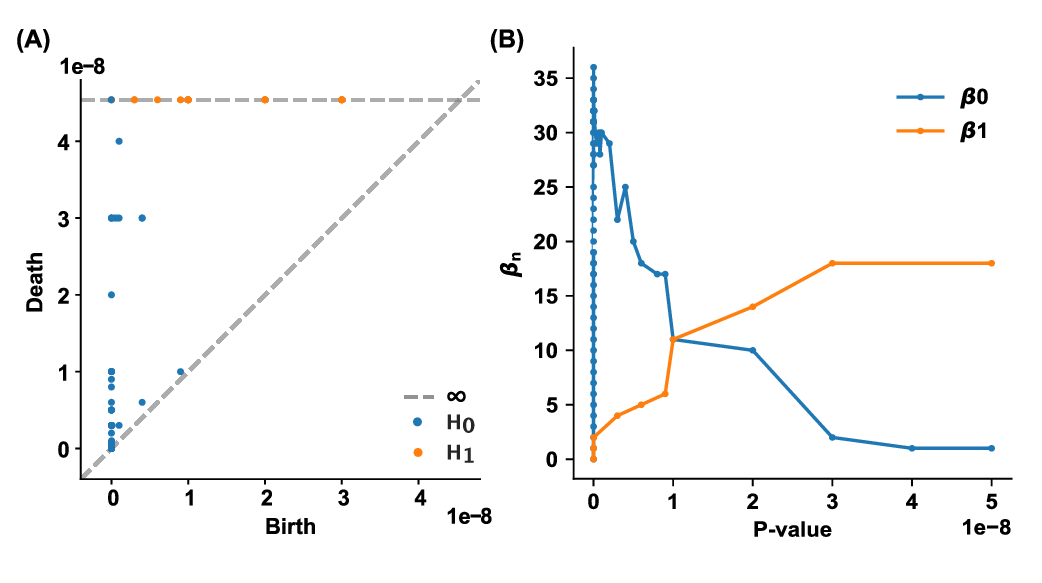}
\caption{Persistent homology analysis of the T2D GWAS hits in the protein–protein interaction network. (A) The persistence diagram of the T2D GWAS hits. Using persistent homology, 0-dimensional holes (connected components, marked as blue dots) and 1-dimensional holes (loops, marked as orange dots) were identified as a function of the P-value threshold. The birth and death pairs of topological features are shown. (B) The Betti numbers of the simplicial homology groups as a function of the P-value threshold. T2D, type 2 diabetes; GWAS, genome-wide association study.}
\label{fig:fig2}
\end{figure}

We examined how the topological features of the T2D disease module evolve and persist in the PPI network as the P-value threshold increases from 0 to 5×10$^{-8}$. In our framework, each node appears at the P-value assigned to that gene. We used the P-value of each T2D GWAS hit as a varying threshold and determined the timing of the appearance (birth) and disappearance (death) of $n$-dimensional holes at different threshold values. In the 0th homology group ($H_0$) analysis, 61 0-dimensional holes (connected components) were identified, of which only one persisted over all P-value thresholds (Figure \ref{fig:fig2}A). This persistent 0-dimensional hole is the LCC, that is, the T2D observable disease module. In the 1st homology group ($H_1$) analysis, 18 1-dimensional holes (loops or 1-cycles) were identified, all of which persisted over all P-value thresholds (Figure \ref{fig:fig2}A). No higher-dimensional hole ($n\ge2$) existed in the T2D GWAS hit data. The Betti numbers (ranks) of the simplicial homology groups are shown in Figure \ref{fig:fig2}B. As the P-value threshold increases, the number of 0-dimensional holes converges to 1 (i.e., the LCC), while the number of 1-dimensional holes increases and converges to 18.

\begin{figure}[h]
\centering
\includegraphics[width=0.85\linewidth]{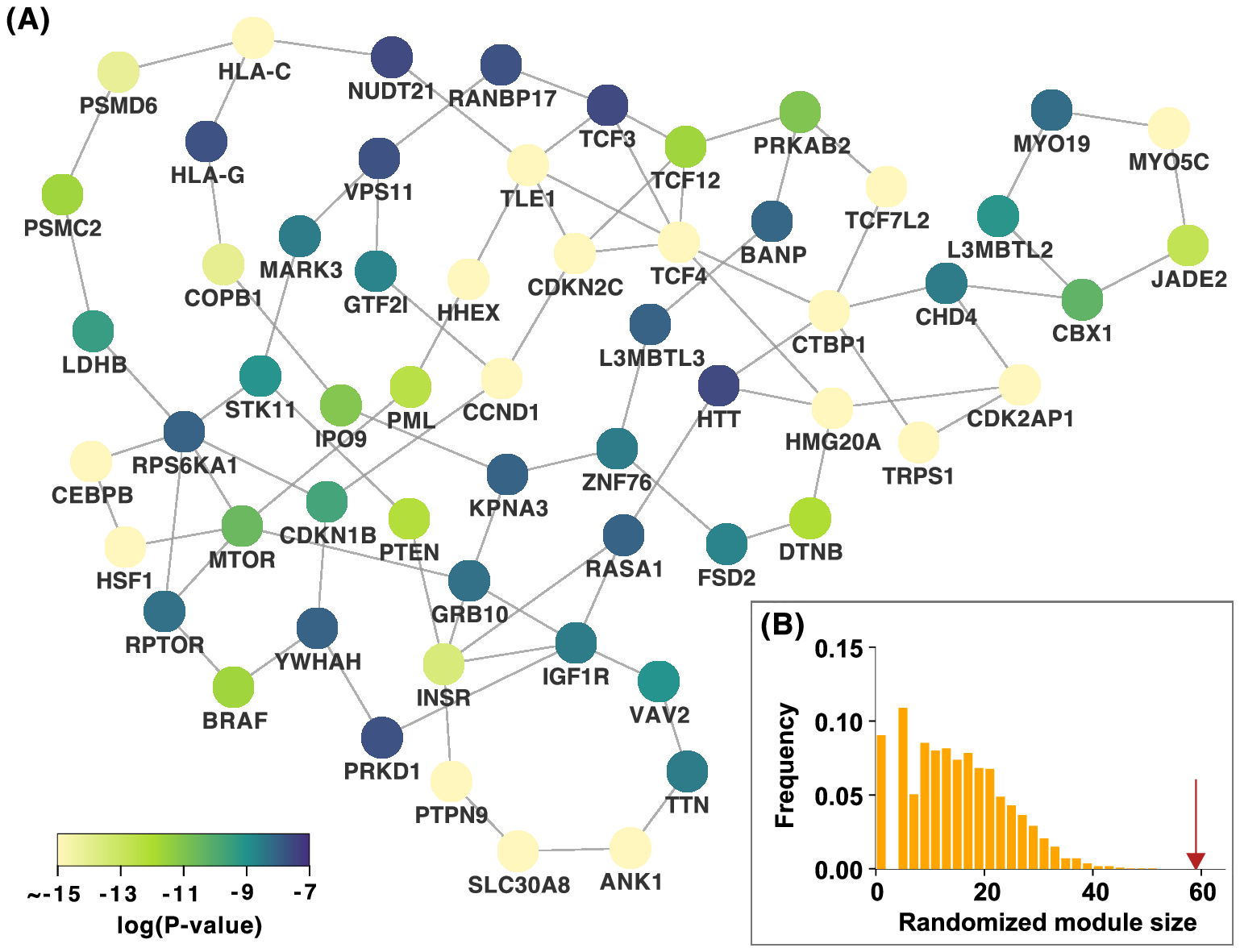}
\caption{The 1st persistent T2D disease module in the protein–protein interaction network. (A) The 1st persistent T2D disease module comprising persistent 1-dimensional holes (loops) that live forever over all P-value thresholds. The P-values of the T2D GWAS hits are shown. (B) The 1st persistent T2D disease module is significantly larger than that expected by chance. The red arrow indicates the observed module size. T2D, type 2 diabetes; GWAS, genome-wide association study.}
\label{fig:fig3}
\end{figure}

We identified the $n$-th persistent disease modules, which were defined as unions of persistent $n$-dimensional holes that live forever over all P-value thresholds. The 0th persistent disease module is the LCC, which is the persistent 0-dimensional hole that lives forever. As shown in Figure \ref{fig:fig1}B, the LCC is significantly larger than that expected by chance. Since the LCC concept has been extensively investigated in various complex diseases, this study focused on the 1st persistent disease module. In our T2D GWAS data analysis, we identified 18 persistent 1-dimensional holes (loops or 1-cycles), which constitute the 1st persistent T2D disease module comprising 59 nodes and 83 edges (Figure \ref{fig:fig3}A). This 1st persistent T2D disease module is significantly larger than that expected by chance (P$<$0.001, Figure \ref{fig:fig3}B), indicating a significant topological feature of the T2D GWAS hits in the PPI network. Since the smallest P-value in the T2D GWAS data is extremely small as 3e-695 (rs7903146 in \textit{TCF7L2}), we repeated our analysis using the log P-value scale. The same 61 0-dimensional holes and 18 1-dimensional holes were also identified (Supplementary Figure S1).

\subsection{Biological pathways, transcription factors, and microRNAs}
\begin{figure}[h]
\centering
\includegraphics[width=0.9\linewidth]{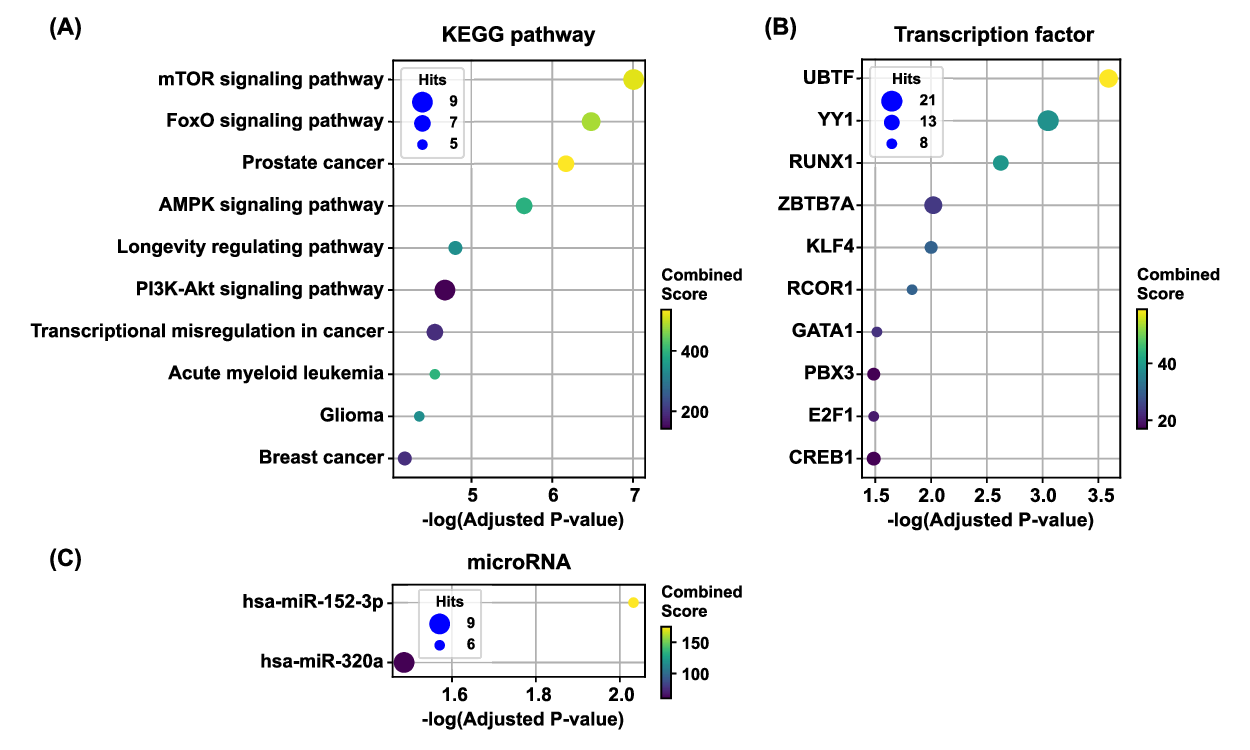}
\caption{Functional enrichment analysis of the 1st persistent T2D disease module. The 1st persistent T2D disease module was enriched by KEGG pathways (the top 10 pathways are shown) (A), targets of transcription factors (B), and targets of microRNAs (C). T2D, type 2 diabetes; KEGG, the Kyoto Encyclopedia of Genes and Genomes.}
\label{fig:fig4}
\end{figure}

To infer the pathobiological significance of the 1st persistent T2D disease module, we identified over-represented KEGG pathways. The top 10 enriched KEGG pathways included mTOR signaling, FoxO signaling, AMPK signaling, the longevity regulating pathway, PI3K-Akt signaling, the transcriptional misregulation pathway in cancer, and several cancer pathways (Figure \ref{fig:fig4}A). In addition, the 1st persistent T2D disease module was enriched with targets of transcription factors, including UBTF, YY1, RUNX1, ZBTB7A, KLF4, RCOR1, GATA1, PBX3, E2F1, and CREB1 (Figure \ref{fig:fig4}B). The 1st persistent T2D disease module was also enriched with targets of microRNAs, including hsa-miR-152-3p and hsa-miR-320a (Figure \ref{fig:fig4}C).

\section{Discussion}
Using persistent homology, this study explored the evolution and persistence of the topological features of T2D GWAS hits in the PPI network as the P-value threshold increased from 0 to 5×10$^{-8}$. The $n$-th persistent disease module was defined as a union of persistent $n$-dimensional holes that live forever over all P-value thresholds. This is a higher-order generalization of the conventional disease module concept. The 0th persistent T2D disease module is the LCC of the T2D GWAS hits, which is significantly larger than that expected by chance. In the 1st homology group analysis, all 18 1-dimensional holes (loops) of the T2D GWAS hits persist over all P-value thresholds. The 1st persistent T2D disease module comprising these 18 persistent 1-dimensional holes is significantly larger than that expected by chance, indicating a significant topological structure in the PPI network. The 1st persistent T2D disease module is enriched with the mTOR, FoxO, AMPK, and PI3K-Akt signaling pathways; longevity regulating pathway; and cancer pathways. It has been known that the mechanisms of T2D, aging, and cancer are closely related to each other \cite{Wei2017}. The pathobiological significance of this persistent disease module is subject to subsequent experimental validation.

Our computational topology framework potentially has broad applicability to other complex phenotypes. By analyzing the persistent homology features, the higher-order topological features that may be closely associated with a specific phenotype can be identified. We plan to expand this preliminary study to systematically analyze the topological features of the large-scale disease–gene networks \cite{Menche2015, Guney2016}. We expect that there are several mathematical ways to expand our persistent homology approach in PPI networks. The weighted topology \cite{Baccini2022} of weighted PPI networks reflecting proteome-wide binding affinity and concentration information should provide more biologically plausible and reliable information. In addition, relational persistent homology \cite{Stolz2023} may be a useful tool for dissecting multispecies data, such as multiomics data or multilayer biological networks.

Notwithstanding, this study has several potential limitations. While the conventional disease module concept typically relies on connected components (0th homology class) of disease seeds, the proposed persistent disease module concept is a higher-order generalization of the LCC. Hence, it is hard to directly compare our homology approach to most other disease module identification algorithms. Therefore, It is essential to develop higher-order versions of seed-expanding algorithms to detect robust and reliable persistent disease modules. The role of seed connectors \cite{Wang2018} in homology features also warrants elucidation. As the uncertainty and incompleteness of GWAS and PPI network data are inevitable \cite{Menche2015}, how these errors and uncertainty affect the robustness of persistent disease modules remains unclear. Although no higher-dimensional hole ($n\ge2$) was present in our T2D GWAS hit data, higher-order interactions may play a significant role in disease pathobiology. Dynamic topological data analysis approaches based on sequential data would provide more rigorous and robust results \cite{Ciocanel2021}. Tissue- or cell-type-specific networks \cite{Greene2015} should provide more biological information regarding persistent disease modules. Determining whether oncogenic mutations perturb PPI or higher-order interactions in the PPI network is a worthwhile endeavor \cite{Cheng2021}.

\section*{Acknowledgement}
This study received no external funding. The author has no conflicts of interest to declare. The author would like to thank Dr. Ruisheng Wang for kindly sharing the consolidated human protein–protein interactome data and the anonymous reviewers for their valuable comments.

\section*{Authorship Contribution}
\textbf{Euijun Song:} Conceptualization, Methodology, Formal analysis, Investigation, Software, Visualization, Writing – original draft.

\bibliographystyle{apalike}
\bibliography{refs}

\end{document}